# QUANTUM INTERFERENCES IN THE RAMAN CROSS SECTION FOR THE RADIAL BREATHING MODE IN METALIC CARBON NANOTUBES

G. Bussi[1], J. Menéndez[2], J. Ren[2], M. Canonico[2], and E. Molinari[1]

[1] INFM National Research Center on nanoStructures and bioSystems at Surfaces (S3) and Dipartimento di Fisica, Università di Modena e Reggio Emilia, Via Campi 213/A, I-41100 Modena, Italy

[2] Department of Physics and Astronomy, Arizona State University, Tempe, AZ 85287-1504

**ABSTRACT**

The lineshapes of the Raman excitation profiles for radial breathing modes in carbon nanotubes are shown to be strongly affected by interference effects that arise whenever strong optical transitions are separated by a small energy. This is the case in metallic zig-zag and chiral tubes, where one-dimensional singularities in the electronic joint density of states are split due to the trigonal warping of the electronic band structure of two-dimensional graphene. It is shown that the proper modeling of these interferences is crucial for the identification of the ($n,m$) indices using Raman spectroscopy.

The spectroscopy of one-dimensional Van Hove singularities in the electronic density of states is critical for the characterization of carbon nanotubes and for the understanding of their electronic structure. Several techniques can be used to study these singularities, including optical absorption [1-3], photoluminescence[4, 5], photoluminescence excitation spectroscopy [6], Resonance Raman Scattering (RRS) [7-13], and scanning tunneling spectroscopy [14-16]. Unfortunately, the nanotube species ($n,m$) (we use the standard notation to identify nanotubes, see Ref. [17]) present in a sample are usually not known *a priori* because no precise control over the indices $n$ and $m$ can be achieved with current fabrication techniques. As a result of this limitation, the assignment of the observed electronic transitions to specific ($n,m$) structures is very challenging. Since electronic singularities detected with RRS by the radial breathing mode (RBM) can be assigned to a specific nanotube diameter [7], RRS has become a major tool for narrowing down the possible ($n,m$) choices for a given set of experimental data. A fundamental implicit assumption in these studies is the existence of a one-to-one correspondence between maxima in the resonant excitation profiles and singularities in the electronic joint density of states. This is equivalent to neglecting the contribution to the scattering amplitude of all but the electronic states closest to the singularity in resonance, a reasonable expectation given the divergent nature of one-dimensional critical points. In fact, detailed calculations for armchair ($n,n$) tubes have fully confirmed this picture [18]. In this letter, however, we show that for realistic electronic structure parameters strong interferences are predicted between contributions from nearby singularities in metallic chiral and zigzag tubes. These interferences distort the excitation profiles to the extent that accurate critical point energies cannot be extracted from the experimental data without detailed modeling.

For the analysis of Raman scattering in a single carbon nanotube it is convenient to assume an infinite array of parallel tubes, separated by a distance such that the tube-tube interactions are negligible. Let us consider incident(scattered) light with frequency $\omega_L$ ($\omega_S$), wave vector $K_L$ ($K_S$) and polarization $\eta_L$ ($\eta_S$). The Raman scattering cross section is given by

$$\frac{d\sigma}{d\Omega} = \left(\frac{n_L}{\omega_L}\right)\left(\frac{\omega_S n_S}{c^4}\right)\left(\frac{V}{2\pi\hbar}\right)^2 \left|W_{fi}(\omega_S, K_S, \eta_S; \omega_L, K_L, \eta_L)\right|^2 \quad (1)$$

where $V$ is the volume of the sample, $c$ the speed of light in vacuum, $n(\omega_L)$ [$n(\omega_S)$] the index of refraction at frequency $\omega_L$ [$\omega_S$], and $W_{fi}$ the quantum-mechanical transition matrix element between the initial and final states. The collection angle element $d\Omega$ is in the direction of $K_S$. The measured Raman cross section as a function of the laser photon frequency $\omega_L$ is the so-called Raman excitation profile (REP). Richter and Subbaswamy [18] were the first to apply the conventional theory of RRS in crystalline solids to the calculation of $W_{fi}$ in carbon nanotubes. Analytical expressions were derived by Canonico *et al.* (Ref. [13]), neglecting excitonic effects[19, 20] and exploiting the fact that the electronic energy bands near a singularity have a parabolic dispersion. The results of Ref. [13] can be rewritten as:

$$W_{fi} \propto \sum_t C_t \left[\frac{1}{(\hbar\omega_L - E_t - i\Gamma_t)^{1/2}} - \frac{1}{(\hbar\omega_S - E_t - i\Gamma_t)^{1/2}}\right], \quad (2)$$

with

$$C_t = |P_z^t|^2 (m_t^*)^{1/2} \frac{dE_t}{dR}, \qquad (3)$$

where $P_z^t$ is the $z$-component ($z$ is taken along the nanotube axis) of the momentum matrix element between the bands involved in the optical transition $t$, $m_t^*$ the reduced mass of the virtually excited electron-hole pair, and $dE_t/dR$ is the derivative of the singularity's transition energy relative to the radius of the nanotube.

When $\hbar\omega_L$ approaches the energy of a transition $E_{jj}$ in an armchair tube with a diameter within the experimental range, good agreement is obtained between the numerical calculations of Richter and Subbaswamy [18] and the results of Eq. (2) with a single $t = jj$ term in the summation. This confirms that REPs in these armchair tubes can be understood "one singularity at a time." The reason for this behavior is that interband optical transitions in armchair tubes with realistic diameters are well separated. On the other hand, if two or more singularities are sufficiently close, their contributions to Eq. (2) will produce noticeable interferences when Eq. (2) is inserted in Eq. (1). Moreover, since the prefactor $C_t$ is different for each singularity, the shape of REP will depend very sensitively on the relative magnitudes *and signs* of these prefactors for each of the contributing transitions.

In order to investigate interference effects in real systems, we consider metallic chiral and zigzag nanotubes. We expect excitonic effects to be of lesser importance in these systems, so that a theory based on free electron-hole pairs is more likely to be valid. More importantly, metallic chiral and zigzag nanotubes are strong candidates for interference effects due to the splitting of their van Hove singularities caused by trigonal warping in the band structure of two-dimensional graphene [21].

In Figure 1, we show the electronic band structure for a (15,0) tube. The bands were obtained from an *ab initio* density-functional theory (DFT) calculation within the local density approximation (LDA). We use norm-conserving pseudopotentials and a plane-wave basis set [22] with an energy cutoff at 50 Rydberg and 6 *k*-points along the tube direction. A supercell with a 35 Å width is used to mimic isolated tubes. Since we work within a plane-wave formalism, it is straightforward to compute the matrix elements of the momentum operator for each possible interband optical transition [23]. Based on this analysis we identify the lowest two strong nearby optical transitions, labeled $E_{11}^-$ (= 1.99 eV) and $E_{11}^+$ (=1.62 eV) which are split by trigonal warping. For the Raman cross section, we use Eq. (2) with two terms, corresponding to transitions $E_{11}^-$ and $E_{11}^+$, and insert the resulting transition matrix element into Eq. (1). For the computation of $C_{11}^-$ and $C_{11}^+$ we need the momentum matrix elements for each transition, $|P_z(11-)|^2 = 0.32$ Å$^{-2}$ and $|P_z(11+)|^2 = 0.75$ Å$^{-2}$, and the reduced electron-hole mass, which we obtain by fitting parabolic expressions to the $E(k)$ curves near the bottom of the bands. We find $m^*(11-) = 1.71 m_e$ and $m^*(11+) = 0.4 m_e$, where $m_e$ is the electron mass. The derivative of the transition energy relative to the nanotube radius, $dE_t/dR$, is obtained from separate DFT-LDA calculations at different values of the tube radius, *i.e.* in a frozen description of the RBM. The results are shown in the inset to Fig 1. We find that $dE_t/dR$ has a different sign for the two transitions. In Fig. 2(*a*) we show the predicted REP for the (15,0) tube as a solid line. The dotted line shows the same calculation with the sign of $C_{11}^+$ reversed. If there were no interferences between the two transitions, the two curves would be identical. This is clearly not the case, demonstrating that interference effects cannot be neglected. Moreover, since the separation between the split singularities is maximum for zigzag tubes [21], we expect *weaker* interferences in such systems and *stronger* interferences in metallic chiral tubes. This is schematically illustrated in Figure 2(*b*), where we repeat the calculations with the separation between $E_{11}^-$ and $E_{11}^+$ artificially reduced to 50% of its value for the (15,0) tube.

It is apparent from Fig. 2 and Eq. (3) that the relative sign of $dE_t/dR$ for the two trigonal-warping-split transitions has a dramatic impact on the predicted REPs. We find that the opposite sign we obtain for the (15,0) tube is actually a common feature of all metallic chiral and zigzag nanotubes. This can be understood in terms of the simple graphene model of the electronic structure of carbon nanotubes. The electronic bands along the $\Gamma$–$K$ direction of the graphene Brillouin zone are approximately given by $\varepsilon = \pm (t + 2t \cos x)$, where $x$ is a dimensionless wave vector such that $x_K = 2\pi/3$ and $t$ is the nearest neighbor hopping integral. Therefore, the energy of the corresponding optical transitions is $E(x) = 2 |t + 2t \cos x|$. For a metallic ($n$,0) zigzag tube, the transitions $E_{jj}^+$ and $E_{jj}^-$ correspond to graphene interband transitions at $x = x_K \pm j\pi/n$. Note that conduction and valence bands are exchanged in the two cases. Since the dispersion curves are not symmetric around $x_K = 2\pi/3$ (trigonal warping), $E_{jj}^- > E_{jj}^+$, *i.e.*, the two transitions are not degenerate. If the graphene sheet is distorted following the displacement pattern for a RBM, the bands become $E = \pm (t_2 + 2t_1 \cos x)$ [24], where $t_2 = t$ and $t_1$ oscillates as a function of the instantaneous C-C separation. Let us suppose for example that $t_1$ becomes slightly smaller than $t_2$. Then the band dispersion is reduced, so that one the bands moves toward $+ t_2$ while the other one moves toward $- t_2$. As a result of this flattening, $E_{jj}^+ = 2|t_1 + 2t_2 \cos (x_K + j\pi/n)|$ decreases and $E_{jj}^- = 2|t_1 + 2t_2 \cos (x_K - j\pi/n)|$ increases for $j < n/6$. Inci-dentally, a tight-binding fit of the pseudopotential results, assuming $t_1 = t(d_0/d)^n$, where $d$ is the interatomic separation and $d_0$ its value for graphene, gives $n = 3.85$. It is interesting to note that this value is quite different from the well-known $n = 2$ prescription from Harrison[25]. An inverse-cube law has been proposed for carbon systems [26].

If the nanotube electronic structure could be calculated from first principles with an accuracy of a few meV, a single experimental transition energy would suffice for an unambiguous determination of the chiral indices ($n,m$). State-of-the-art electronic structure methods are far from the required accuracy, but this limitation can be overcome—at least partially—by measuring and modeling more than one optical transition in the same nanotube. In particular, the trigonal warping splitting of singularities is very sensitive to chirality because it ranges from zero for armchair tubes to a maximum for zigzag tubes [21]. Thus the *separation* between energy levels is crucial for structural assignments, and it is precisely the experimental determination of this separation which is strongly affected by the interference effects discussed above. In Fig 3 we show a REP corresponding to a 188 cm$^{-1}$ RBM, measured by Canonico *et al.*[13]. This profile deviates for the single-peak curve expected for armchair tubes, so it was assigned to a chiral or zigzag tube. From the relationship between tube diameter and RBM frequency the possible candidates are (12,6), (11,8), (16,1), (15,3), (14,5) and (13,7) tubes. We first fit the REP as a superposition of two transitions by adding two Lorentzian shapes, as frequently done in the literature. We obtain $E_{11}^- = 1.890$ eV and $E_{11}^+ = 1.795$ eV., i.e., a separation of 95 meV. However, if we fit the same REP using Eq. (2) and constraining the ratio $C_{11}^-/C_{11}^+$ to be negative, as indicated by theory, our best fit gives $E_{11}^- = 1.910$ eV, $E_{11}^+ = 1.790$ eV. The separation between the two levels is now 120 meV, a 25% increase. Furthermore, a fit of comparable quality can be obtained by using Eq. (2) and $C_{11}^-/C_{11}^+ > 0$, but in this case we obtain $E_{11}^- = 1.870$ eV and $E_{11}^+ = 1.799$ eV, that is, the separation between the energy levels is reduced to 79 meV.

We now show that the strong dependence of the fitted energy separation on the REP profile assumption leads to different (*n,m*) assignments, underscoring the importance of properly treating interference effects in structural determinations based on Raman spectroscopy. We use the simple nearest-neighbor tight-binding model introduced above, with a single adjustable parameter *t*. We adjust this parameter to fit the one of the two experimental transition energies, for example $E_{11}^+$, for the six candidate nanotubes. We obtain values between *t* = 2.735 eV and *t* = 3.02 eV, which are all within the known uncertainty of this parameter. In other words, the calculation of a single transition energy does not lead to a narrowing of structural assignments for the REP in Fig. 3. However, once the parameter *t* is fixed, the simple tight-binding theory makes a prediction for the separation of the trigonal-warping split levels. This prediction is sensitive to the tube's chirality. Moreover, since we are fitting one of the transitions exactly (and we know that the tight binding model reproduces quite well the general band structure shape) we expect the predicted $E_{11}^-$ - $E_{11}^+$ separation to be more reliable than the absolute value of the individual energies. The calculated values for $E_{11}^-$ - $E_{11}^+$ are 110 meV (12,6); 52 meV (11,8); 210 meV (16,1); 180 meV (15,3); 140 meV (14,5); and 92 meV (13,7). Therefore, the best assignment is (12,6) if we use Eq. 2 to model the REP lineshape. Had we chosen the customary Lorentzian model, however, (or Eq. 2 with $C_{11}^-/C_{11}^+ > 0$) we might have concluded that the best assignment for the 188 cm$^{-1}$ REP is a (13,7) tube. Incidentally, for $C_{11}^-/C_{11}^+ < 0$ the fit value for $C_{11}^-/C_{11}^+$ is –2.22, which is remarkably close to the theoretical prediction $C_{11}^-/C_{11}^+ = -1.86$ for a (15,0) tube with a 5% smaller diameter.

In summary, we have shown that interferences between different electronic transitions have a profound and unexpected effect on the Raman excitation profiles in carbon nanotubes. Neglecting such interferences may lead to systematic errors in the estimates of optical transition energies from Raman experiments. For the analysis of experimental REPs, as well as for the use of Stokes/anti-Stokes ratios to determine transition energies, it is crucial to use realistic theoretical models if one expects to obtain energies with the required precision to discriminate between different (*n*, *m*) values. The calculations and comparisons with experiment presented here are limited to the RBM, but the effect might be expected to be even more significant for the totally symmetric tangential modes, since the ratio between phonon energy and singularity splitting will be larger for these modes.

We are grateful to E. Chang and A. Ruini for useful discussions. This work was funded by the National Science Foundation under grant NSF-DMR 0244290 and by INFM through a CINECA supercomputing project. The support by the RTN EU Contract ``EXCITING'' No. HPRN-CT-2002-00317, and by FIRB ``NOMADE'' is also acknowledged.

**FIGURE CAPTIONS**

**Figure 1**  Electronic band structure of a zigzag (15,0) tube near the Fermi level. The arrows indicate the two transitions $E_{11}^-$ and $E_{11}^+$ that have a sizable oscillator strength. The inset shows the dependence of the energy of these transitions on the radius of the nanotube, and the vertical dotted line indicates the reference radius.

**Figure 2**  (a) Solid line: Predicted Raman excitation profile for the RBM in a (15,0) zigzag nanotube using Eqs. (2) and (1) and electronic parameters as discussed in the text. In addition, we assumed a phonon energy $\hbar(\omega_L - \omega_S) = 0.022$ eV and a broadening of $\hbar\Gamma = 0.07$ eV. Dotted line: same calculation as in the case of the solid line, but with the sign of $C_{11A}/C_{11B}$ reversed. The vertical lines indicate the energies of the two singularities.
(b) Same as part (a) but with the energy separation $E_{11}^- - E_{11}^+$ artificially reduced to one-half of its original value.

**Figure 3**  Experimental REP for a 188 cm$^{-1}$ RBM (from Ref. [13]) and a fit (solid line) using Eqs. (2) and (1) with the same broadening parameter for the two transitions. The fit values are $E_{11}^- = 1.910$ eV, $E_{11}^+ = 1.790$ eV, $C_{11}^-/C_{11}^+ = -2.22$, and $\Gamma = 0.045$ eV.

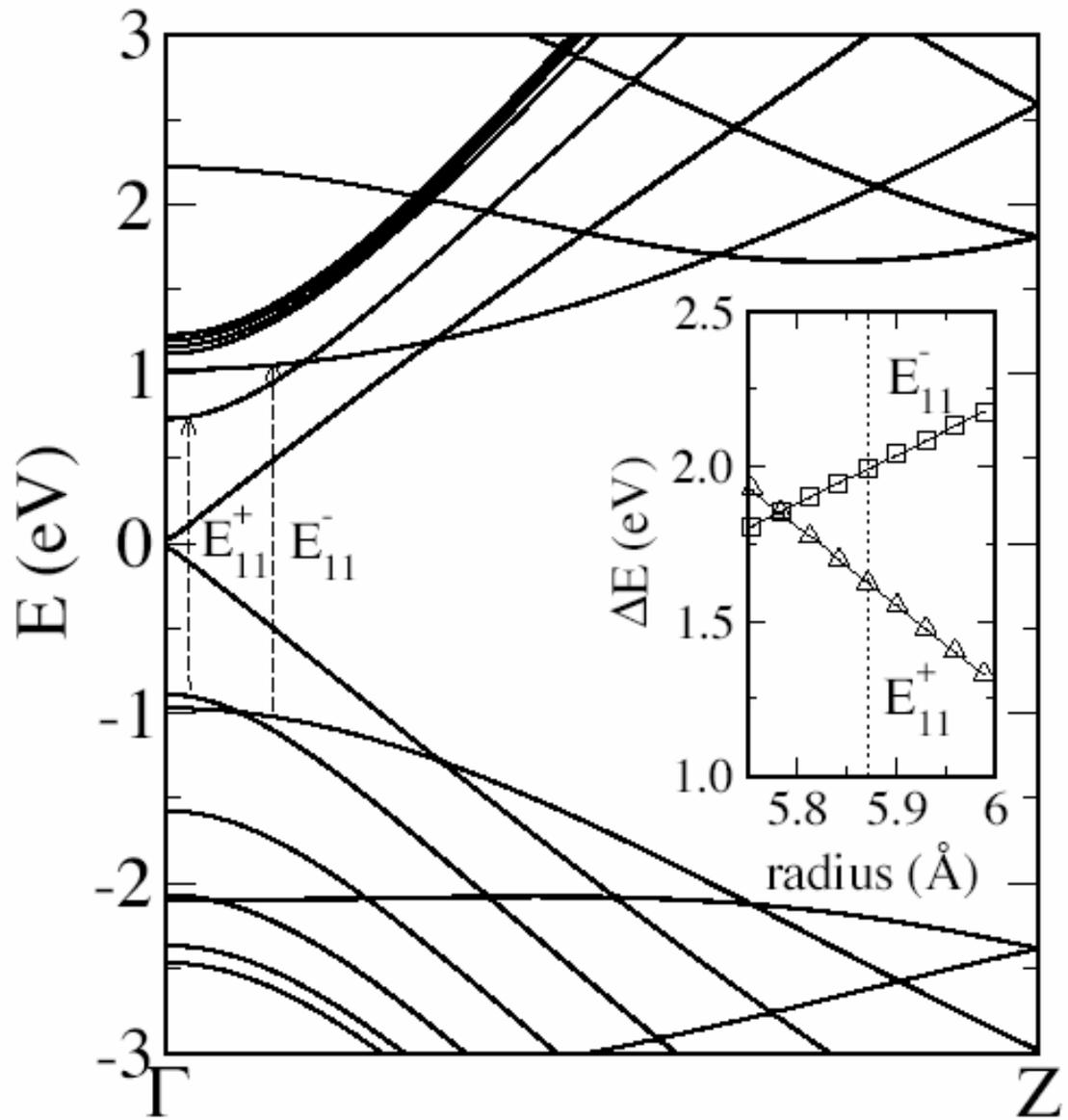

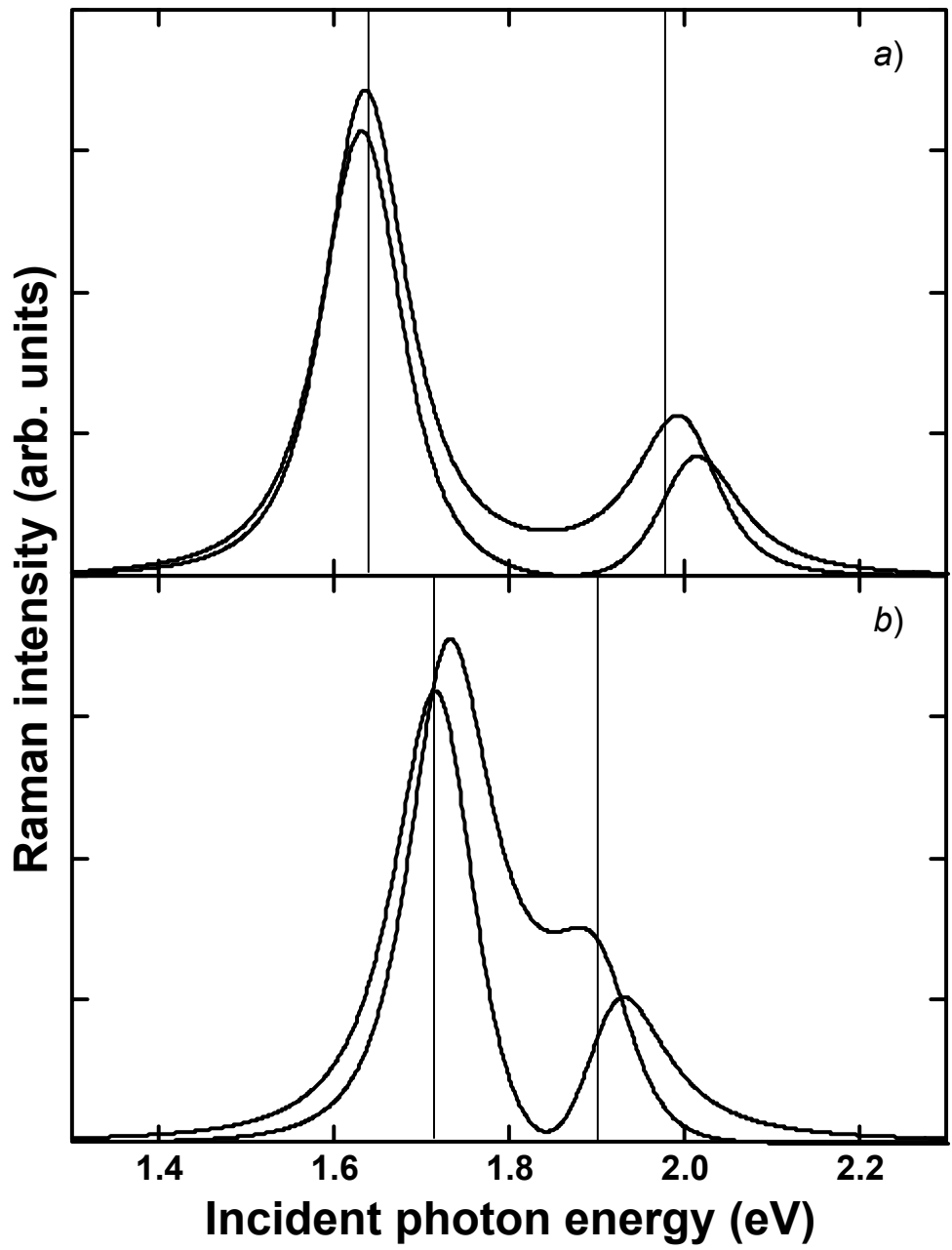

**Figure 2**

**Bussi et al.**

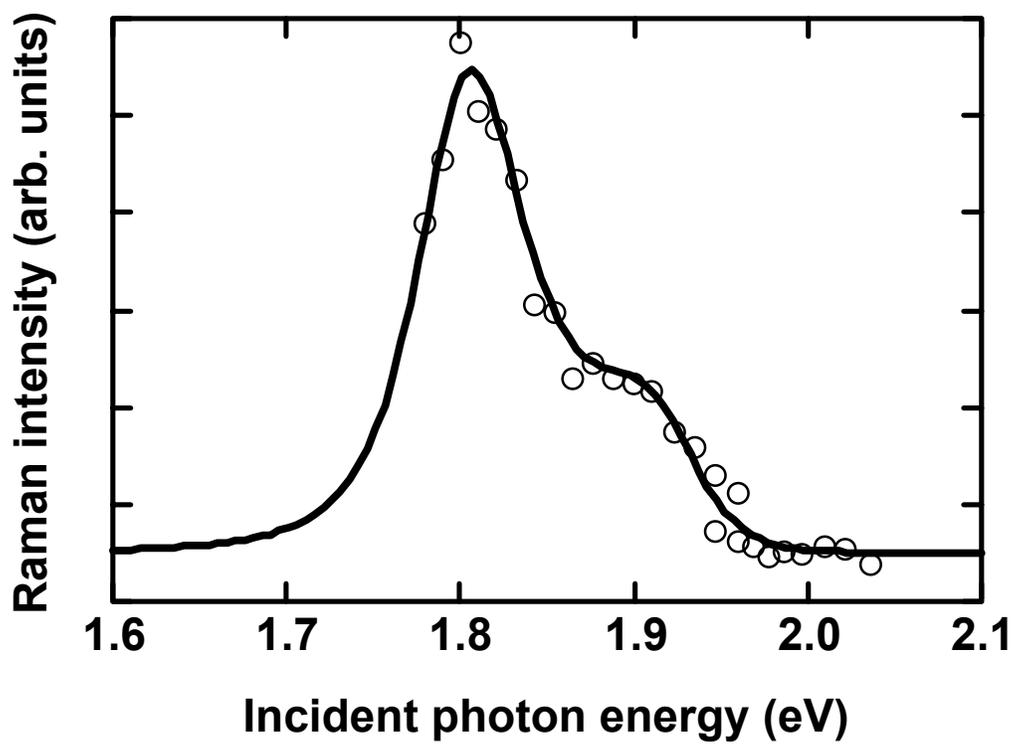

**Figure 3**

**Bussi et al.**